\documentclass[aps,prd,groupedaddress,showpacs,superscriptaddress]{revtex4-1}
\usepackage{graphicx}
\usepackage{dcolumn}
\usepackage{bm}
\begin{document}

\title {Vortex dynamics in nonrelativistic Abelian Higgs model.}

\author{A.~A.~Kozhevnikov}
\email[]{kozhev@math.nsc.ru} \affiliation{Laboratory of
Theoretical Physics, S.~L.~Sobolev Institute for Mathematics, Novosibirsk, Russian
Federation}
\affiliation{Novosibirsk State University, Novosibirsk, Russian
Federation}

\date{\today}

\begin{abstract}
The dynamics of the gauge vortex with arbitrary form of a contour
is considered in the framework of the nonrelativistic Abelian
Higgs model,   including the possibility of the gauge field
interaction with the fermion asymmetric background. The equations
for the time derivatives of the curvature and the torsion of the
vortex contour  generalizing the Betchov-Da Rios equations in
hydrodynamics, are obtained. They are applied to study the
conservation of helicity of the gauge field forming the vortex,
twist, and writhe numbers  of the vortex contour. It is shown that
the conservation of helicity is broken when both  terms in the
equation of the vortex motion  are present, first due to the
exchange of excitations of the phase and modulus of the scalar
field and the second one due to the coupling of the gauge field
forming the vortex, with the fermion asymmetric background.
\end{abstract}

\maketitle

{\bf 1. Introduction.} The string-like (or vortex) solutions of
the field-theoretical models are widely discussed in application to
numerous physical systems, from the really  observed in condensed
matter physics (quantized vortices in He$^4$ \cite{Donn},
Abrikosov lines in type II superconductors \cite{Abrikos},
vortices in Bose - Einstein condensates \cite{Frbec} etc.) to
hypothetical cosmic strings \cite{Schellard}. Of particular
interest are the situations when the dynamics of the vortex
contours can be deduced from the field equations of the underlying
field theory. Different type of models are invoked to study the
dynamics. In particular, the dynamics of cosmic strings was
studied in the framework of the Nambu-Goto action
\cite{NG,Scherk}. The latter  was shown to result
\cite{Forst} from the relativistic version of the Abelian Higgs
model (AHM). The solution in the form of the static cosmic string
was obtained from the static version of AHM \cite{ANO}. Another
limiting case is the nonrelativistic one, which can be deduced
from  the nonrelativistic Abelian Higgs model (NRAHM). The static
limit of this model coincides with the Ginzburg-Landau theory
\cite{GL}.

The vortex in the models with the local gauge symmetry is called
the local one. The equation of the nonrelativistic  motion  of the
local vortex, in the idealized situation of the negligible
dissipation, was obtained in  the work \cite{Ark10} in the
framework of NRAHM. The feature of the equation of the vortex
motion in the model considered in Ref.~\cite{Ark10} is that except
for the usual term stating that the vortex ring velocity points to
the direction of bi-normal, there appears the additional term
arising due to the exchange of the excitations of the phase and
modulus of the scalar field between different segments of the
contour. Yet the model considered in \cite{Ark10} is not the most
general one. As was pointed out in \cite{Ark99}, there could be
another term in the gauge vortex equation of motion in case when
the gauge field is coupled to the asymmetric background of chiral
fermions \cite{red,rub,joyce,laine,semik}. In the static case, the
influence of this additional term on the form of the  vortex
contour was considered in Ref.~\cite{Ark99}.

Another interesting aspect of the time-dependent gauge vortex
solutions is that they can elucidate the interrelations among such
characteristics of the closed contours as the helicity of gauge
field forming the vortex, the writhe and torsion (twist) numbers
and their possible dependence on time \cite{moff,berg,ric}. In
case of pure hydrodynamics, the corresponding equation was deduced
in Ref.~\cite{darios}. Note that the problem of dynamical
evolution of the vortex filaments has a long story traced to the
beginning of the twentieth century \cite{darios}. The
corresponding equations for the time derivatives of the curvature
and torsion were rediscovered, in particular, in
Refs.~\cite{betch,hasim}. The historical review of these
developments is presented in Ref.~\cite{ricnat}.

The aim of the present work is to consider the dynamical evolution
of the gauge vortex string in NRAHM, including the possibility of
its interaction with the static fermion asymmetric background.

{\bf 2. Nonrelativistic Abelian Higgs Model with the gauge
vortex.} Nonrelativistic Abelian Higgs model incorporating gauge
vortices is given by the following Lagrangian density
\cite{Ark10}:
\begin{eqnarray}
{\cal L}&=&\frac{1}{8\pi}({\bm E}^2-{\bm
H}^2)-\frac{g}{2}(|\psi|^2-n_0)^2+\frac{1}{2}[\psi^\ast(i\hbar\partial_t-q\varphi+qa_0)\psi
+{\rm
c.c.}]-\nonumber\\&&\frac{1}{2m}\left|\left(-i\hbar{\bm\nabla}-\frac{q}{c}{\bm
A}+\frac{q}{c}{\bm a}\right)\psi\right|^2-\rho_0\varphi,
\label{L}\end{eqnarray} where  ${\bm E}=-\partial_t{\bm
A}/c-{\bm\nabla}\varphi$ and ${\bm H}={\bm\nabla\times{\bm A}}$
are the electric and magnetic field strengths,  the four-vector
gauge potential is  $A_\mu=(\varphi,{\bm A})$; $q$, $m$ are the
charge, mass, of the particles forming the condensate of the
scalar field, $c$ is the velocity of light. The quantities $n_0$
and $\rho_0$ are, respectively, the density of the scalar field
condensate  and the homogenous positive charge density introduced
to provide the net neutrality of the system:
\begin{equation}
\rho_0+qn_0=0.\label{neutrality}\end{equation} The coupling
constant $g$ can be related  to the sound velocity $c_s$. See
Eq.~(\ref{cs}) below. The limiting case of this model, with the
gauge coupling constant set to zero, corresponds to the model of
Gross-Pitaevskii \cite{Pit,Gro} possessing the solutions in the
form of global vortices observed in Bose systems like HeII or BEC.
The vortex is represented as the line of singular phase $\chi_s$
of the scalar field $\psi=\sqrt{n_0}e^{i\chi_s}$ which forms the
spatially homogeneous condensate of the density $n_0$ everywhere
except for the vortex core, where it goes to zero at the
transverse distances of the order of the healing (or correlation)
length $\xi$.

The notations in Eq.~(\ref{L}) are as follows. The four-vector
$$a_\mu=-\frac{\hbar c}{q}\partial_\mu\chi_s$$ is the
four-gradient of the singular phase,
\begin{equation}
[{\bm\nabla}\times{\bm\nabla}]\chi_{\rm s}=2\pi\int d\sigma{\bm
X}^\prime\delta^{(3)}({\bm x}-{\bm
X}),\label{chising}\end{equation} serving as the source of the
gauge vortex with the unit flux quantum whose space-time location
is given by the vector ${\bm X}\equiv{\bm X}(t,\sigma)$. The
length of the contour, $l=\int_{\sigma_{i}}^{\sigma_{f}}|{\bm
X}^\prime|d\sigma$, is the natural choice of $\sigma$. The
explicit expressions for $a_\mu$ found from Eq.~(\ref{chising}),
in the gauge ${\bm\nabla}{\bm a}=0$, are:
\begin{eqnarray}\label{amu}
a_0&=&\frac{\hbar}{2q}\oint\frac{([\dot{{\bm X}}\times{\bm X}^\prime],{\bm x}-{\bm X})}{|{\bm x}-{\bm X}|^3}d\sigma, \nonumber\\
{\bm a}&=&\frac{\hbar c}{2q}\oint\frac{[{\bm X}^\prime\times({\bm
x}-{\bm X})]}{|{\bm x}-{\bm X}|^3}d\sigma.\label{amu}
\end{eqnarray}
Hereafter, the prime (dot) means the differentiation with respect
to the contour parameter $\sigma$ (time), and $[{\bm a}\times{\bm
b}]$ [$({\bm a},{\bm b})$] stands for the vector (scalar) product
of two vectors ${\bm a}$ and ${\bm b}$.

The term $\propto g$ in Eq.~(\ref{L}) can be rewritten in the form
$$-\frac{g}{2}(|\psi|^2-n_0)^2=-\frac{g}{2}|\psi|^4+gn_0|\psi|^2-\frac{g}{2}n^2_0,$$
demonstrating that the first term in the right hand side describes the s-wave
interaction between the bosons forming the condensate, the second
one is $\mu|\psi|^2$, with the chemical potential $\mu=gn_0$
pertinent for the Bose gas with the s-wave interaction \cite{LP},
and the third one is irrelevant constant. After integration over
space and going to the Hamiltonian this second term goes to $-\mu
N$ thus showing that the actual Hamiltonian is in fact
$H^\prime=H-\mu N$ \cite{LP}. This chemical potential $\mu$ is due
to the background of the bosonic condensate. The equation of motion of
the gauge vortex resulting from NRAHM was shown \cite{Ark10} to
look like
\begin{eqnarray}
[\dot{{\bm X}}\times{\bm X}^\prime]&=&\frac{\hbar}{2m}\left({\bm
X}^{\prime\prime}\ln\frac{\lambda_L}{\xi}+\frac{1}{c^2_s}\left[\left(\frac{\partial}{\partial
t}[\dot{{\bm X}}\times{\bm X}^\prime]\right)\times{\bm
X}^\prime\right]\ln\frac{\lambda_s}{\xi}\right),\label{loceqmo}
\end{eqnarray}where
\begin{equation}
\lambda_L=\sqrt{\frac{mc^2}{4\pi
n_0q^2}}\label{lambdaL}\end{equation} is the the London penetration depth,
\begin{equation}
c_s=\sqrt{\frac{n_0g}{m}}.\label{cs}\end{equation} is the velocity of sound \cite{Bog},
\begin{equation}
\xi=\frac{\hbar}{2mc_s}\label{xi}\end{equation}is the healing
length (the characteristic size of the vortex core).  The term
$\propto 1/c^2_s$  and the intermediate scale
\begin{equation}
\lambda_s=\lambda_L\frac{c_s}{c}\ll\lambda_L,\label{lambs}
\end{equation}
as was shown in Ref.~\cite{Ark10}, are  generated dynamically, due
to the exchange of the fluctuations of the phase and the modulus
of the scalar field $\psi$ between the different segments of the
gauge vortex. The  term $[\dot{{\bm X}}\times{\bm X}^\prime]$
originates from  the variation of the term $\int
q|\psi|^2a_0dtd^3x$ in the action \cite{Ark10}. For further use
let us introduce also the quantum of magnetic flux
\begin{equation}\label{Phi0}
\Phi_0=\frac{2\pi\hbar c}{q}.\end{equation}

It is known \cite{red,rub,laine} that the presence of the
background of chiral fermions (whose asymmetry is characterized by the chemical potential $\mu_f$)
coupled to the gauge field $A_\mu$ (in particular, the Abelian one
\cite{laine,semik}), induces the term $-\mu_fN_{\rm CS}$ in the
Hamiltonian, where
\begin{equation}
N_{\rm CS}=\frac{1}{\Phi^2_0}\int d^3x({\bm
A},[{\bm\nabla}\times{\bm A}])\label{NCS}\end{equation}is the
Chern-Simons number carried by the Abelian gauge field of the
vortex. Coupling of the Abelian gauge field with the chiral
fermion \cite{laine} takes place, in particular, in the standard
model of particle physics \cite{joyce} where the right electron is
coupled to the hypermagnetic gauge field. As is seen from
Eq.~(\ref{NCS}), the Chern-Simons number in the Abelian case is
proportional to helicity,
\begin{equation}
h_A=\int d^3x({\bm A},[{\bm\nabla}\times{\bm
A}]).\label{hA}\end{equation}Using the expressions for  ${\bm A}$
and ${\bm B}=[{\bm\nabla}\times{\bm A}]$ from Ref.~\cite{Ark10},
\begin{eqnarray}\label{AB}
{\bm A}({\bm
x},t)&=&\Phi_0\int\frac{d^3k}{(2\pi)^3}\frac{i/\lambda^2_L}{{\bm
k}^2({\bm k}^2+1/\lambda^2_L)}\times\oint d\sigma[{\bm
k}\times{\bm
X}^\prime] e^{i({\bm k},{\bm x}-{\bm X})},\nonumber\\
{\bm H}({\bm
x},t)&=&\Phi_0\int\frac{d^3k}{(2\pi)^3}\frac{1/\lambda^2_L}{{\bm
k}^2+1/\lambda^2_L}\times\oint d\sigma{\bm X}^\prime e^{i({\bm k},
{\bm x}-{\bm X})},
\end{eqnarray}
one obtains the expression for the helicity of the single vortex
with the unit flux quantum and its variation over ${\bm X}$:
\begin{widetext}
\begin{eqnarray}\label{hAF}
h_A&=&i\Phi_0^2\oint
d\sigma_1d\sigma_2\int\frac{d^3k}{(2\pi)^3{\bm
k}^2}\left(\frac{1/\lambda^2_L}{{\bm
k}^2+1/\lambda^2_L}\right)^2\left({\bm k},\left[\frac{\partial{\bm
X}(\sigma_1)}{\partial\sigma_1}\times\frac{\partial{\bm
X}(\sigma_2)}{\partial\sigma_2}\right]\right)\exp\left[-i({\bm
k},{\bm X}(\sigma_1)-{\bm X}(\sigma_2)\right],\nonumber\\
\delta
h_A&=&2\Phi_0^2\int\frac{d^3k}{(2\pi)^3}\left(\frac{1/\lambda^2_L}{{\bm
k}^2+1/\lambda^2_L}\right)^2\oint d\sigma_1d\sigma_2(\delta{\bm
X}_1,[{\bm X}^\prime_1\times{\bm X}^\prime_2])e^{-i({\bm k},{\bm
X}_{12})}=\nonumber\\&&\frac{\Phi_0^2}{4\pi\lambda^3_L}\oint
d\sigma_1d\sigma_2(\delta{\bm X}_1,[{\bm X}^\prime_1\times{\bm
X}^\prime_2])e^{-|{\bm X}_{12}|/\lambda_L},
\end{eqnarray}
\end{widetext}
where ${\bm X}_{1,2}\equiv{\bm X}(\sigma_{1,2})$, ${\bm
X}_{12}={\bm X}_1-{\bm X}_2$. Since the segments of the vortex
contour located at the distances greater than $\lambda_L$ give
exponentially damped contribution, one can use the expansion in
$z=\sigma_2-\sigma_1$,
\begin{equation}\label{zexpan}
{\bm X}(\sigma+z)={\bm X}(\sigma)+z{\bm X}^\prime(\sigma)+\frac{z^2}{2}{\bm X}^{\prime\prime}(\sigma)+\frac{z^3}{6}{\bm X}^{\prime\prime\prime}(\sigma)+...
\end{equation}
and keep only the first non-vanishing term surviving in the limit
$\lambda_L\to\infty$:
\begin{eqnarray}\label{delhA1}
\delta h_A&=&\frac{\Phi_0^2}{8\pi\lambda^3_L}\oint d\sigma([{\bm
X}^\prime\times{\bm X}^{\prime\prime\prime}],\delta{\bm
X})\int_{-\infty}^\infty
z^2e^{-|z|/\lambda_L}dz=\frac{\Phi_0^2}{2\pi}\oint d\sigma([{\bm
X}^\prime\times{\bm X}^{\prime\prime\prime}],\delta{\bm X}).
\end{eqnarray}The variation of the corresponding term in the action,
$\Delta S=\mu_f\int dtN_{\rm CS}$, gives the term
$\propto[{\bm X}^\prime\times{\bm X}^{\prime\prime\prime}]$ in the equation of motion \cite{Ark99} which now looks like
\begin{widetext}
\begin{eqnarray}
[\dot{{\bm X}}\times{\bm X}^\prime]&=&\gamma{\bm
X}^{\prime\prime}+T_0\left[\left(\frac{\partial}{\partial
t}[\dot{{\bm X}}\times{\bm X}^\prime]\right)\times{\bm
X}^\prime\right]+\tilde{\mu}[{\bm X}^\prime\times{\bm
X}^{\prime\prime\prime}],\label{toteqmo1}
\end{eqnarray}\end{widetext}
where the following shorthand notations are used:
\begin{eqnarray}
\gamma&=&\frac{\hbar}{2m}\ln\frac{\lambda_L}{\xi},\nonumber\\
T_0&=&\frac{\hbar}{2mc^2_s}\ln\frac{\lambda_s}{\xi},\nonumber\\
\tilde{\mu}&=&\frac{\mu_f}{4\pi^2\hbar
n_0}.\label{not1}\end{eqnarray} The term $\gamma{\bm
X}^{\prime\prime}$ is due to the tension which push the vortex
segment to the direction of the normal vector ${\bm n}$. Other
terms are discussed below. When both $T_0$ and $\tilde{\mu}$ are
set to zero, the known relation $\dot{{\bm X}}=\gamma\kappa{\bm
b}$ is recovered. In this case the static solution with zero
velocity is possible for the straight vortex for which $\kappa=0$,
$\tau=0$. The particular dynamical solution to the latter equation
describes the vortex ring moving with the velocity
$V=\gamma\kappa$ in the direction orthogonal to the ring plane. If
$T_0=0$, $\tilde{\mu}\not=0$ then the nontrivial static solution
in the form of the helix with  $\kappa=$const and
$\tau=\gamma/\tilde{\mu}$ is possible \cite{Ark99}.

Taking into account the fact that the physical motion of the
vortex is orthogonal to the unit vector ${\bm X}^\prime$, one can
rewrite Eq.~(\ref{toteqmo1}) in the form of the wave equation with
the additional terms:
\begin{equation}\label{toteqmo2}
\frac{1}{c^2_0}\ddot{{\bm X}}-{\bm X}^{\prime\prime}+\frac{1}{\gamma}
[\dot{{\bm X}}\times{\bm X}^\prime]+\frac{\tilde{\mu}}{\gamma}[{\bm X}^\prime\times{\bm
X}^{\prime\prime\prime}]=0,
\end{equation}where
\begin{equation}\label{c0}
  c_0=\left(\frac{\gamma}{T_0}\right)^{1/2}=c_s\left(\frac{\ln\lambda_L/\xi}{\ln\lambda_s/\xi}\right)^{1/2}.
\end{equation}The interpretation of the additional terms is the following.
The term $\propto[\dot{{\bm X}}\times{\bm X}^\prime]$ is the
analog of the Magnus force acting on the moving vortex due to the
nonzero circulation of the supercurrent. In the classical
hydrodynamics, in the direction orthogonal to the displacement
velocity, the velocity of the circulating flow adds with the
displacement velocity on one side, and subtracts from it on the
opposite side. By Bernoulli's equation, the pressure in the liquid
is greater in the regions where the velocity is lower, yielding
the force in the direction orthogonal to the vectors $\dot{{\bm
X}}$ and ${\bm X}^\prime$. Of course, in the case of our interest
the situation is much more intricate because one should take into
account the role of a number of quantum excitations, but the
qualitative picture is probably the same.

The term $\propto[{\bm X}^\prime\times{\bm
X}^{\prime\prime\prime}]$ is induced by the anomaly. Indeed, the
variation of the induced Chern-Simons term in the action over
vector potential ${\bm A}$ results in the anomalous current ${\bm
j}_{\rm an}\propto\mu_f{\bm H}$ pointed along the direction of the
magnetic field. This relation is the corner stone of the widely
discussed chiral magnetic effect. See the review \cite{kharz} and
references therein. This current is subjected to the action of the
Lorentz force $\propto[{\bm j}_{\rm an}\times{\bm H}]$ from the
magnetic field produced by the nearby segments of the gauge
vortex. Taking into account the exponential damping of the
magnetic field strength and the expansion (\ref{zexpan}), one can
write the leading contribution to the anomalous part of the
Lorentz force per unit length:
\begin{equation}\label{fl}
{\bm f}_L\propto\mu_f\int_{-\infty}^\infty dz[{\bm
X}^\prime(\sigma)\times{\bm
X}^\prime(\sigma+z)]e^{-|z|/\lambda_L}\propto\mu_f[{\bm
X}^\prime\times{\bm X}^{\prime\prime\prime}].
\end{equation} So,
the discussed term is the macroscopic manifestation of the quantum
anomaly. The helical form of the vortex contour in the static case
\cite{Ark99} is due to the balance between the tension and the
anomaly-induced Lorentz force.

{\bf 3. Time derivatives of the  basic contour characteristics.} As is
known, the derivatives over the contour parameter $\sigma$ of the
basic unit contour vectors of normal ${\bm n}$, bi-normal ${\bm
b}$, and tangent ${\bm X}^\prime$ constituting the right triple,
${\bm X}^\prime=[{\bm n}\times{\bm b}]$, are given by the Frenet-Serret  equations
\begin{equation}
\frac{\partial}{\partial\sigma}\left(%
\begin{array}{c}
  {\bm X}^\prime \\
  {\bm n} \\
  {\bm b} \\
\end{array}%
\right)=\left(%
\begin{array}{ccc}
  0 & \kappa & 0 \\
  -\kappa & 0 & \tau \\
  0 & -\tau & 0 \\
\end{array}%
\right)\left(%
\begin{array}{c}
  {\bm X}^\prime \\
  {\bm n} \\
  {\bm b} \\
\end{array}%
\right),\label{Frenet}\end{equation}where $\kappa$, $\tau$ are the
curvature and torsion, respectively. One can write down the
analogous set of equations for the time derivatives of the above
vectors taking into account the fact that they are unit ones and
that the physical motions of the contour are transverse (i.e.
orthogonal to ${\bm X}^\prime$):
\begin{equation}
\frac{\partial}{\partial t}\left(%
\begin{array}{c}
  {\bm X}^\prime \\
  {\bm n} \\
  {\bm b} \\
\end{array}%
\right)=\left(%
\begin{array}{ccc}
  0 & a_{Xn} & a_{Xb} \\
  a_{nX} & 0 & a_{nb} \\
  a_{bX} & a_{bn} & 0 \\
\end{array}%
\right)\left(%
\begin{array}{c}
  {\bm X}^\prime \\
  {\bm n} \\
  {\bm b} \\
\end{array}%
\right)\equiv\hat{A}\left(%
\begin{array}{c}
  {\bm X}^\prime \\
  {\bm n} \\
  {\bm b} \\
\end{array}%
\right)\label{tder}.\end{equation}Equating the commutator of the
derivatives over contour parameter and time to zero one finds that
the matrix $\hat{A}$ in (\ref{tder}) is antisymmetric,
$a_{bn}=-a_{nb}$, $a_{nX}=-a_{Xn}$, and the following equations
are valid:
\begin{eqnarray}
a_{nb}&=&(\tau a_{Xn}+a^\prime_{Xb})/\kappa ,\nonumber\\
\dot{\kappa}&=&a^\prime_{Xn}-\tau a_{Xb},\nonumber\\
\dot{\tau}&=&\kappa
a_{Xb}+a^\prime_{nb},\label{genrel}\end{eqnarray} so that
\begin{equation}
\hat{A}=\left(%
\begin{array}{ccc}
  0 & a_{Xn} & a_{Xb} \\
  -a_{Xn} & 0 & a_{nb} \\
  -a_{Xb} & -a_{nb} & 0 \\
\end{array}%
\right)\label{A1}\end{equation}One can relate $\dot{\kappa}$ and
$\dot{\tau}$ with the components of the vortex velocity $\dot{{\bm
X}}$, or, equivalently, with the dual vector ${\bm W}=[\dot{{\bm
X}}\times{\bm X}^\prime]=W_n{\bm n}+W_b{\bm b}$, $\dot{{\bm
X}}=W_n{\bm b}-W_b{\bm n}$. Differentiating the definition of the
vector ${\bm W}$ over $\sigma$ and using the Frenet-Serret
equations and Eq.~(\ref{tder}), one obtains
\begin{equation}\label{Xdotpr}
\frac{\partial{\bm X}^\prime}{\partial t}={\bm b}(W^\prime_n-\tau
W_b)-{\bm n}(W^\prime_b+\tau W_n).
\end{equation}
This equation permits one to make the identification
\begin{eqnarray}\label{ais}
a_{Xn}&=&-W^\prime_b-\tau W_n,\nonumber\\
a_{Xb}&=&W^\prime_n-\tau W_b,
\end{eqnarray}
so that
\begin{eqnarray}\label{Darios2}
\dot{\kappa}&=&-(W^\prime_b+\tau W_n)^\prime-\tau(W^\prime_n-\tau
W_b),\nonumber\\
\dot{\tau}&=&\kappa(W^\prime_n-\tau
W_b)+\left\{\left[(W^\prime_n-\tau W_b)^\prime-\tau(W^\prime_b+
\tau W_n)\right]/\kappa\right\}^\prime.
\end{eqnarray}
With these ingredients one can rewrite Eq.~(\ref{toteqmo1}) in
terms of the components of ${\bm W}$:
\begin{eqnarray}\label{eqmo2}
W_n&=&\kappa(\gamma-\tilde{\mu}\tau)+T_0\left\{\dot{W}_b+\left[(W^\prime_n-\tau W_b)^\prime-\tau(W^\prime_b+\tau W_n)\right]W_n/\kappa\right\},\nonumber\\
W_b&=&\tilde{\mu}\kappa^\prime-T_0\left\{\dot{W}_n-\left[(W^\prime_n-\tau
W_b)^\prime-\tau(W^\prime_b+\tau W_n)\right]W_b/\kappa\right\}.
\end{eqnarray}The system of equations (\ref{Darios2}) is in fact the result
of definitions which are not related to the specific dynamical
model. The latter is provided by Eq.~(\ref{eqmo2}). In particular,
when $T_0$ and $\tilde{\mu}$ both vanish, $W_n=\gamma\kappa$,
$W_b=0$, one finds in this limiting case that
\begin{eqnarray}\label{Darios0}
\dot{\kappa}&=&-\gamma(2\kappa^\prime\tau+\kappa\tau^\prime),\nonumber\\
\dot{\tau}&=&\gamma\left[\kappa\kappa^\prime+\left(\frac{\kappa^{\prime\prime}}{\kappa}-\tau^2\right)^\prime\right].
\end{eqnarray}
After  the re-scaling $\kappa\to\gamma^{-1/2}\kappa$,
$\tau\to\gamma^{-1/2}\tau$, $\sigma\to\gamma^{1/2}\sigma$ the
parameter $\gamma$ drops from equations, and  one gets the Betchov-Da Rios equations \cite{darios,betch,hasim} known in
hydrodynamics for a long time.

Equations (\ref{eqmo2}) are  nonlinear, so let us find their
solution which is exact in the fermion chemical potential
$\tilde{\mu}$ but perturbative  to the first order in $T_0$. The
zeroth order solution is
\begin{eqnarray}\label{Ws0}
W^{(0)}_n&=&\kappa(\gamma-\tilde{\mu}\tau),\nonumber\\
W^{(0)}_b&=&\tilde{\mu}\kappa^\prime.
\end{eqnarray}The corresponding analog of the Betchov-Da Rios equations found from (\ref{Darios2}) looks as
\begin{eqnarray}\label{betchmu0}
\dot{\kappa}^{(0)}&=&-(\gamma-\tilde{\mu}\tau)(2\kappa^\prime\tau+\kappa\tau^\prime)-\tilde{\mu}(\kappa^{\prime\prime}-\kappa\tau^2),\nonumber\\
\dot{\tau}^{(0)}&=&\kappa[\gamma\kappa^\prime-\tilde{\mu}(2\kappa^\prime\tau+\kappa\tau^\prime)]
+\left\{\left[(\gamma-\tilde{\mu}\tau)(\kappa^{\prime\prime}-\kappa\tau^2)+
\tilde{\mu}(2\kappa^\prime\tau+\kappa\tau^\prime)^\prime\right]/\kappa\right\}^\prime
\end{eqnarray}It is straightforward to obtain the first order correction $\propto T_0$ using the expressions \begin{widetext}
\begin{eqnarray}\label{pert1}
W^{(1)}_n&=&T_0\left\{\dot{W}_b+\frac{1}{\kappa}\left[(W^\prime_n-\tau
W_b)^\prime-\tau(W^\prime_b+\tau
W_n)\right]W_n\right\}^{(0)},\nonumber\\
W^{(1)}_b&=&-T_0\left\{\dot{W}_n-\frac{1}{\kappa}\left[(W^\prime_n-\tau
W_b)^\prime-\tau(W^\prime_b+\tau W_n)\right]W_b\right\}^{(0)},
\end{eqnarray}\end{widetext}
but the corresponding expressions are excessively long. Instead,
one can find the time derivative of the total torsion. From the
third equation in (\ref{genrel}) and the second equation in
(\ref{ais}) one obtains
\begin{equation}\label{dotT0}
\oint\dot{\tau}d\sigma=\oint\kappa
a_{Xb}d\sigma=-\oint(\kappa^\prime W_n+\kappa\tau
W_b)d\sigma.\end{equation}Inserting here Eqs.~(\ref{Ws0}) and
(\ref{pert1}) and using Eq.~(\ref{betchmu0}), after some lengthy
algebraic manipulations and integrations by parts, one finds
\begin{equation}\label{dotT}
\oint\dot{\tau}d\sigma=\frac{1}{4}\gamma
T_0\tilde{\mu}\oint\kappa^4\tau^\prime d\sigma.\end{equation}This
expression shows that the total torsion of the gauge vortex in the
nonrelativistic Abelian Higgs model  is not conserved in case of
the presence of coupling of the gauge field forming the vortex,
with the fermion asymmetric background described by
$\tilde{\mu}\not=0$. Coming back to the physical quantities one
obtains
\begin{equation}\label{dotT1}
\oint\dot{\tau}d\sigma=\frac{\hbar\mu_f}{64\pi^2m^2c^2_sn_0}\times\ln\frac{\lambda_L}{\xi}\times\ln\frac{\lambda_s}{\xi}\oint\kappa^4\tau^\prime
d\sigma.\end{equation}One should have in mind that the equation of
motion (\ref{loceqmo}) was obtained in the paper \cite{Ark10} in
the London limit $\lambda_L/\xi\gg1$, $\lambda_s/\xi\gg1$, so
Eq.~(\ref{dotT1}) tells us that the time derivative of the total
torsion, being proportional to the small factor $1/c^2_s$, is
small but not negligible, because it is multiplied by two large
logarithmic factors. Note also that the coherence length $\xi$
characterizes the radius of the vortex core where the condensate
of the scalar field vanishes.

The time derivative of helicity is obtained from
Eq.~(\ref{delhA1}):
\begin{widetext}
\begin{eqnarray}\label{dothA}
\dot{h}_A&=&\frac{\Phi_0^2}{2\pi}\oint d\sigma([\dot{{\bm
X}}\times{\bm X}^\prime],{\bm
X}^{\prime\prime\prime})=\frac{\Phi_0^2}{2\pi}\oint d\sigma({\bm
W},\kappa^\prime{\bm n}+\kappa\tau{\bm
b})=\frac{\Phi_0^2}{2\pi}\oint d\sigma(\kappa^\prime
W_n+\kappa\tau
W_b)=-\frac{\Phi_0^2}{2\pi}\oint\dot{\tau}d\sigma\equiv\nonumber\\&&-\Phi_0^2\frac{d{\rm Tw}}{dt},
\end{eqnarray}\end{widetext}where ${\rm Tw}$ is the twist number,
\begin{equation}\label{Tw}
{\rm Tw}=\frac{1}{2\pi}\oint\tau d\sigma=\frac{1}{2\pi}\oint({\bm X}^\prime,[{\bm n}^\prime\times{\bm n}])d\sigma.\end{equation}

One can relate the above quantities with the writhe number. The
term $([{\bm X}^\prime\times{\bm
X}^{\prime\prime\prime}],\delta{\bm X})$  in the integrand of
expression for the variation of helicity (\ref{delhA1}), hence the
corresponding  term $([{\bm X}^\prime\times{\dot{\bm X}}],{\bm
X}^{\prime\prime\prime})$ in the expressions for the time
derivatives (\ref{dothA}), has the geometrical meaning. Indeed,
let us find the writhe and its variation:
\begin{widetext}
\begin{eqnarray}
{\rm Wr}&=&\frac{1}{4\pi}\oint d\sigma_1\oint
d\sigma_2 \frac{({\bm X}_{12},[{\bm X}^\prime_1\times{\bm X}^\prime_2])}{|{\bm X}_{12}|^3}=
-i\int\frac{d^3k}{(2\pi)^3}\oint d\sigma_1d\sigma_2\frac{({\bm k},[{\bm X}^\prime_2\times
{\bm X}^\prime_1])}{{\bm k}^2}\times e^{i{\bm k}\cdot{\bm X}_{21}},\nonumber\\
\delta {\rm Wr}&=&-i\int\frac{d^3k}{(2\pi)^3}\frac{{\bm k}}{{\bm
k}^2}\oint d\sigma_1d\sigma_2\left\{[\delta{\bm
X}^\prime_2\times{\bm X}^\prime_1]+[{\bm
X}^\prime_2\times\delta{\bm
X}^\prime_1]+\right.\nonumber\\&&\left.i[{\bm
X}^\prime_2\times{\bm X}^\prime_1]({\bm k}\cdot{\bm
X}_{21})\right\}e^{i{\bm k}\cdot{\bm
X}_{21}}=\int\frac{d^3k}{(2\pi)^3}\frac{{\bm k}}{{\bm k}^2}\oint
d\sigma_1d\sigma_2\left\{-[\delta{\bm X}_2\times{\bm
X}^\prime_1]({\bm k}\cdot{\bm
X}^\prime_2)+\right.\nonumber\\&&\left.[{\bm
X}^\prime_2\times\delta{\bm X}_1]({\bm k}\cdot{\bm
X}^\prime_1)+[{\bm X}^\prime_2\times{\bm X}^\prime_1]({\bm
k}\cdot\delta{\bm X}_2)-[{\bm X}^\prime_2\times{\bm
X}^\prime_1]({\bm k}\cdot\delta{\bm X}_1)\right\}e^{i{\bm
k}\cdot{\bm
X}_{21}}=\nonumber\\&&\int\frac{d^3k}{(2\pi)^3}\frac{{\bm k}}{{\bm
k}^2}\oint d\sigma_1d\sigma_2\left\{[{\bm X}^\prime_2\times[{\bm
k}\times[\delta{\bm X}_1\times{\bm X}^\prime_1]]]-[{\bm
X}^\prime_1\times[{\bm k}\times[{\bm X}_2\times\delta{\bm
X}^\prime_2]]]\right\}\times\nonumber\\&&e^{i{\bm k}\cdot{\bm
X}_{21}}=\int\frac{d^3k}{(2\pi)^3}\oint
d\sigma_1d\sigma_2\left(\delta{\bm X}_{21}\cdot[{\bm
X}^\prime_2\times{\bm X}^\prime_1]\right)e^{i{\bm k}\cdot{\bm
X}_{21}}=\oint d\sigma_1d\sigma_2\left(\delta{\bm
X}_{21}\cdot[{\bm X}^\prime_2\times{\bm
X}^\prime_1]\right)\delta^{(3)}({\bm X}_{21}). \label{dWrF}
\end{eqnarray}\end{widetext}The ambiguity $0\times\infty$ in the integrand can be regularized with the help of expression
$$\delta^{(3)}({\bm
X})=\lim_{\xi\to0}\frac{e^{-{\bm
X}^2/2\xi^2}}{(2\pi)^{3/2}\xi^3}.$$Then, using the expansion
(\ref{zexpan}) in $z=\sigma_2-\sigma_1$, the regularized
expression for the variation of the writhe number can be written
in the form:
\begin{widetext}
\begin{eqnarray}
\delta {\rm Wr}_{\rm
reg}&=&\lim_{\xi\to0}\frac{1}{(2\pi)^{3/2}\xi^3}\oint
d\sigma_1d\sigma_2e^{-{\bm X}^2_{21}/2\xi^2}(\delta{\bm
X}_{21},[{\bm X}^\prime_2\times{\bm
X}^\prime_1])=\lim_{\xi\to0}\frac{1}{(2\pi)^{3/2}2\xi^3}\oint
d\sigma\int_{-\infty}^\infty
dzz^2e^{-z^2/2\xi^2}\times\nonumber\\&&(\delta{\bm X},[{\bm
X}^\prime\times{\bm X}^{\prime\prime\prime}])=\frac{1}{2\pi}\oint
d\sigma(\delta{\bm X},[{\bm X}^\prime\times{\bm
X}^{\prime\prime\prime}])=\delta h_A/\Phi^2_0,\label{dWrreg}
\end{eqnarray}\end{widetext}that is
\begin{equation}\label{dotWr}
\frac{d{\rm
Wr}}{dt}=\frac{\dot{h}_A}{\Phi^2_0}.\end{equation}Combining this
equation and Eq.~(\ref{dothA}) one can see that
\begin{equation}\label{sumcons} \frac{d}{dt}({\rm Wr}+{\rm
Tw})=0,\end{equation} i.e., the sum of the twist and writhe is
conserved as it should \cite{moff,berg,ric,fuller,pohl}. In the
meantime, the helicity of the gauge field configuration in the
model considered here is not conserved. It is essential that the
conservation of helicity is broken due to the fact that  both
terms in the equation of the vortex motion  are present, first due
to the exchange of excitations of the phase and modulus of the
scalar field resulting in the factor $\propto c^{-2}_s$, and the
second one due to the coupling of the gauge field forming the
vortex, with the fermion asymmetric background $\propto\mu_f$. The
latter serves as the source and sink of the helicity. Such
coupling is in fact the consequence of the U(1) anomaly of the
current of the chiral fermions \cite{red}. This anomaly is a quantum
phenomenon hence the appearance of the Planck constant in the
right hand side of Eq.~(\ref{dotT1}).

{\bf 4. Discussion and Conclusion.} The vortex solutions with
nontrivial dynamics in the gauge field models like the
nonrelativistic Abelian one considered here, are interesting in
that they allow to apply some general mathematical concepts like
the topology of curves and the equations for the time derivatives
of the curvature and torsion, to the situations beyond the earlier
considered  hydro- or electrodynamical ones \cite{moff,berg,ric}.
As is shown in the present work, one can relate the above time
derivatives with the vortex velocity. The latter, in turn, is
governed by the equation of motion specific for a given underlying
model. In the presence of both the dynamical correction to the
equation of motion due to exchange of the fluctuations of the
phase and modulus of the scalar field, and the coupling with the
fermion asymmetric background, the helicity of the gauge field is
not conserved. Contrary to the situation in magnetohydrodynamics
where the helicity conservation is violated by the dissipative
effects (finite conductivity of the medium), in the present work,
the nonconservation is reversible, because both signs in the right
hand side of Eq.~(\ref{dotT1}) are possible.

As for the application to physics,  one can hardly hope to observe
the closed or curved vortices in type II superconductors immersed
into the static external magnetic fields. However, their
production could be quite possible in the fast temperature quench
\cite{zur96}. See, for example, Ref.~\cite{zur}, where such a mechanism
was considered in the framework of the  2D Abelian Higgs Model. Also, numerous
simulations of the phase transitions in the early Universe point
to a considerable portion of the closed cosmic strings
\cite{Schellard}. The type of the model considered in the present
work is one where the charged particles interacting with the gauge
field, condense uniformly except for the vortex core. The total
neutrality is provided by some static background of the opposite
charge. The cosmic strings of such type are not excluded in the
models of the dark (hidden) sector \cite{vach09,vach14}. However,
the relation of the present work parameters $q$, $n_0$, and $g$
with ones from the models of the hidden sector, goes beyond the
scope of this paper.

The global vortices, i.e. the vortices in the models with the
global U(1) symmetry, of arbitrary form, were observed in the
dilute Bose or/and Fermi gases. They are described by the limiting
case of the equation of motion (\ref{loceqmo}) in which one should
replace the lengths $\lambda_L$ and $\lambda_s$ by the single
parameter $R$ whose magnitude is of the order of the size of the vessel or the size
of the cloud of the condensed atoms \cite{Ark10}. Of course, the
term $\propto\mu_f$ is excluded in this case because gauge field
is absent, hence $\oint\dot{\tau}d\sigma=0$. In this case, the
torsion and writhe  are conserved separately.


\begin{thebibliography}{99}
\bibitem{Donn}
R.~J.~Donnelly, {\it Quantized Vortices in Helium II}, Cambridge University Press, 1991.
\bibitem{Abrikos}
A.~A.~Abrikosov, ZhETF {\bf32}, 1442 (1957).
\bibitem{Frbec}
K.~W.~Madison, F.~Chavy, W.~Wohlleben, and J.~Dalibard, Phys. Rev. Lett. {\bf84}, 806 (2000).
\bibitem{Schellard}
A.~Vilenkin and E.~P.~S.~Shellard, {\it Cosmic Strings and other Topological Defects}, Cambridge University Press, 1994.
\bibitem{NG}
Y.~Nambu,  Lectures at the Copenhagen Symposium (1970).
\bibitem{Scherk}
J.~Scherk, Rev. Mod. Phys. {\bf47}, 123 (1975).
\bibitem{Forst}
D.~F\"orster, Nucl. Phys. {\bf B81}, (1974) 84.
\bibitem{GL}
V.~L.~Ginzburg and L.~D.~Landau, ZhETF {\bf 20}, 1064 (1950).
\bibitem{ANO}
H.~B.~Nielsen and P.~Olesen, Nucl. Phys. {B61}, 45 (1973).
\bibitem{Ark10}
A.~A.~Kozhevnikov, Int. Journal of Mod. Phys. B{\bf24}, 605 (2010).
\bibitem{Ark99}
A.~A.~Kozhevnikov, Phys. Lett. B{\bf461}, 256 (1999).
\bibitem{red}
A.~N.~Redlich and L.~C.~R.~Wijewardhana, Phys. Rev. Lett. {\bf 54}, 970 (1985).
\bibitem{rub}
V.~A.~Rubakov and A.~N.~Tavkhelidze, Theor. Math. Phys. {\bf65}, 250 (1985); Phys. Lett. B{\bf165}, 109 (1985).
\bibitem{joyce}
M.~Joyce and M.~E.~Shaposhnikov, Phys. Rev. Lett. {\bf79}, 1193 (1997).
\bibitem{laine}
M.~Laine, JHEP 0510:056 (2005).
\bibitem{semik}
V.~B.~Semikoz and J.~W.~F.~Valle, JCAP {\bf 11}, 048 (2011).
\bibitem{kharz}
D.~E.~Kharzeev, Progress in Particle and Nuclear Physics {\bf75}, 133 (2014).
\bibitem{moff}
H.~K.~Moffatt, J. Fluid Mech. {\bf35}, 117 (1969).
\bibitem{berg}
M.~A.~Berger and G.~B.~Field, J. Fluid Mech. {\bf147}, 133 (1984).
\bibitem{ric}
H.~K.~Moffatt and R.~L.~Ricca, Proc. R. Soc. London A{\bf439}, 411 (1992).
\bibitem{darios}
L.~S.~Da Rios, Rend. Circ. Mat. Palermo {\bf22}, 117 (1906).
\bibitem{betch}
R.~Betchov, J. Fluid. Mech. {\bf22}, 471(1965).
\bibitem{hasim}
H.~Hasimoto, J. Fluid. Mech. {\bf51}, 477(1965).
\bibitem{ricnat}
R.~L.~Ricca, Nature {\bf 352}, 561 (1991).
\bibitem{Pit}
L.~P.~Pitaevskii, Sov. Phys. JETP {\bf13}, 451 (1961).
\bibitem{Gro}
E.~P.~Gross, Nuovo Cimento {\bf20}, 454 (1961).
\bibitem{LP}
E.~M.~Lifshits and L.~P.~Pitaevskii, {\it Statistical Physics Part 2}, Nauka Publishers, 1978.
\bibitem{Bog}
N.~N.~Bogolyubov, J. Phys. (USSR) {\bf11}, 23 (1947).
\bibitem{fuller}
F.~B.~Fuller,   Proc. Natl. Acad. Sci. USA {\bf68}, 815, (1971).e
\bibitem{pohl}
W.~F.~Pohl, J. Math. Mech. {\bf17}, 975, (1968).
\bibitem{zur96}
W.~H.~Zurek, Phys. Rep. {\bf276}, 177 (1996).
\bibitem{zur}
G.~J.~Stephens, Lu\'{i}s M.~A.~Bettencourt, and W. H. Zurek, Phys.
Rev. Lett. {\bf88}, 137004 (2002).
\bibitem{vach09}
T.~Vachaspati, Phys. Rev. D{\bf80}, 063502 (2009).
\bibitem{vach14}
J.~M.~Hyde, A.~J.~Long, and T.~Vachaspati, Phys. Rev. D{\bf89}, 065031 (2014).
\end{thebibliography}
\end{document}